\documentclass[11pt]{article}
\usepackage[margin=1in]{geometry}
\usepackage{hyperref}
\usepackage{cite}
\usepackage{amsmath,amssymb,amsfonts}
\usepackage{algorithmic}
\usepackage{graphicx}
\usepackage{textcomp}
\usepackage{xcolor}
\usepackage[capitalise]{cleveref}
\usepackage{threeparttable}
\usepackage{multirow}
\usepackage{array}
\usepackage{booktabs}
\usepackage{authblk}
\usepackage{makecell}   

\def\BibTeX{{\rm B\kern-.05em{\sc i\kern-.025em b}\kern-.08em
		T\kern-.1667em\lower.7ex\hbox{E}\kern-.125emX}}
\newcommand{\keywords}[1]{\par\noindent\textbf{Keywords:} #1\par}

\begin{document}
	\title{Energy-Efficient p-Bit-Based Fully-Connected Quantum-Inspired Simulated Annealer with Dual BRAM Architecture}
	\author[1]{Naoya Onizawa}
	\author[1]{Taiga Kubuta}
	\author[1]{Duckgyu Shin}
	\author[1]{Takahiro Hanyu}
	\affil[1]{Research Institute of Electrical Communication, Tohoku University, Sendai 980-8577, Japan}
	\affil[ ]{Emails: naoya.onizawa.a7@tohoku.ac.jp, ryoma.sasaki.p6@dc.tohoku.ac.jp, duckgyu.shin.p4@dc.tohoku.ac.jp, takahiro.hanyu.c4@tohoku.ac.jp}
	\date{}

\maketitle

\begin{abstract}
Probabilistic bits (p-bits) offer an energy-efficient hardware abstraction for stochastic optimization; however, existing p-bit-based simulated annealing accelerators suffer from poor scalability and limited support for fully connected graphs due to fan-out and memory overhead.
This paper presents an energy-efficient FPGA architecture for stochastic simulated quantum annealing (SSQA) that addresses these challenges. The proposed design combines a spin-serial and replica-parallel update schedule with a dual-BRAM delay-line architecture, enabling scalable support for fully connected Ising models while eliminating fan-out growth in logic resources. By exploiting SSQA, the architecture achieves fast convergence using only final replica states, significantly reducing memory requirements compared to conventional p-bit-based annealers.
Implemented on a Xilinx ZC706 FPGA, the proposed system solves an 800-node MAX-CUT benchmark and achieves up to 50\% reduction in energy consumption and over 90\% reduction in logic resources compared with prior FPGA-based p-bit annealing architectures. These results demonstrate the practicality of quantum-inspired, p-bit-based annealing hardware for large-scale combinatorial optimization under strict energy and resource constraints.
\end{abstract}

	
		\section{INTRODUCTION}
	\label{sec:introduction}

In recent years, probabilistic bits (p-bits) have emerged as a compact hardware abstraction for stochastic inference and optimization \cite{IL}. Leveraging embedded magnetic tunnel junctions \cite{p-bit_device,p-bit_device_fast1,p-bit_device_fast2}, p-bits offer tunable randomness and sub-nanosecond switching, enabling energy-efficient realization of algorithms such as Boltzmann machines \cite{Boltzmann1984}, invertible logic \cite{CIL_training}, Bayesian inference \cite{p-bit_BI}, and Gibbs sampling \cite{p-bit_gibbs}.

A key application of p-bits is simulated annealing (SA) for combinatorial optimization \cite{p-bit_general,TApSA}. Existing stochastic-computing SA accelerators (SSA) implement time-multiplexed arithmetic with FPGA-friendly circuitry \cite{stochastic_first,stochastic,stochastic_book,SSA} but exhibit two critical limitations: (i) spin-parallel datapaths do not scale beyond nearest-neighbor graphs, and (ii) shift-register-based delay lines cause logic and routing cost to grow linearly with the number of spins \cite{JETCAS_SSA,SSQA_FPGA}. These constraints limit the practicality of p-bit-based annealers for dense or fully connected problems despite their algorithmic potential.

This paper introduces a fully connected, energy-efficient stochastic simulated quantum annealing (SSQA) engine that addresses both limitations. The proposed architecture integrates a dual-BRAM delay-line scheme with replica-parallel/spin-serial scheduling to eliminate the $O(N)$ fan-out associated with prior shift-register-based implementations while retaining the proven spin-gate formulation from \cite{SSQA_FPGA}. Centralizing delays in BRAM keeps LUT and FF usage nearly constant as the spin count increases. Implemented on a Xilinx ZC706, the system solves 800-node MAX-CUT benchmarks while sustaining the energy and resource advantages required for embedded deployment.

The contributions of this paper are threefold:
(1) A refined SSQA hardware platform that adapts the existing spin-gate circuit to a spin-serial schedule while preserving fully connected graph support and bounding the per-spin compute cost to $N+1$ cycles;
(2) A novel dual-BRAM delay-line architecture that eliminates the fan-out and routing bottlenecks of prior SSA/SSQA accelerators, leading to flat LUT/FF scaling with the number of spins;
(3) A comprehensive evaluation across hardware prototypes and software baselines, including new sensitivity analyses and application case studies, demonstrating substantial improvements in energy, resource usage, and solution quality.

The remainder of the paper is organized as follows. \cref{sec:preliminary} reviews SSQA and summarizes related annealing hardware. \cref{sec:hardware} details the proposed architecture. \cref{sec:evaluation} reports algorithmic and hardware evaluations, including extended benchmarks. \cref{sec:discussion} analyzes trade-offs and broader implications, and \cref{sec:conclusion} concludes the paper.

\section{PRELIMINARIES}
\label{sec:preliminary}

\subsection{p-bit-based simulated annealing for Ising model}

\begin{figure}[t]
	\centering
	\includegraphics[width=1.0\linewidth]{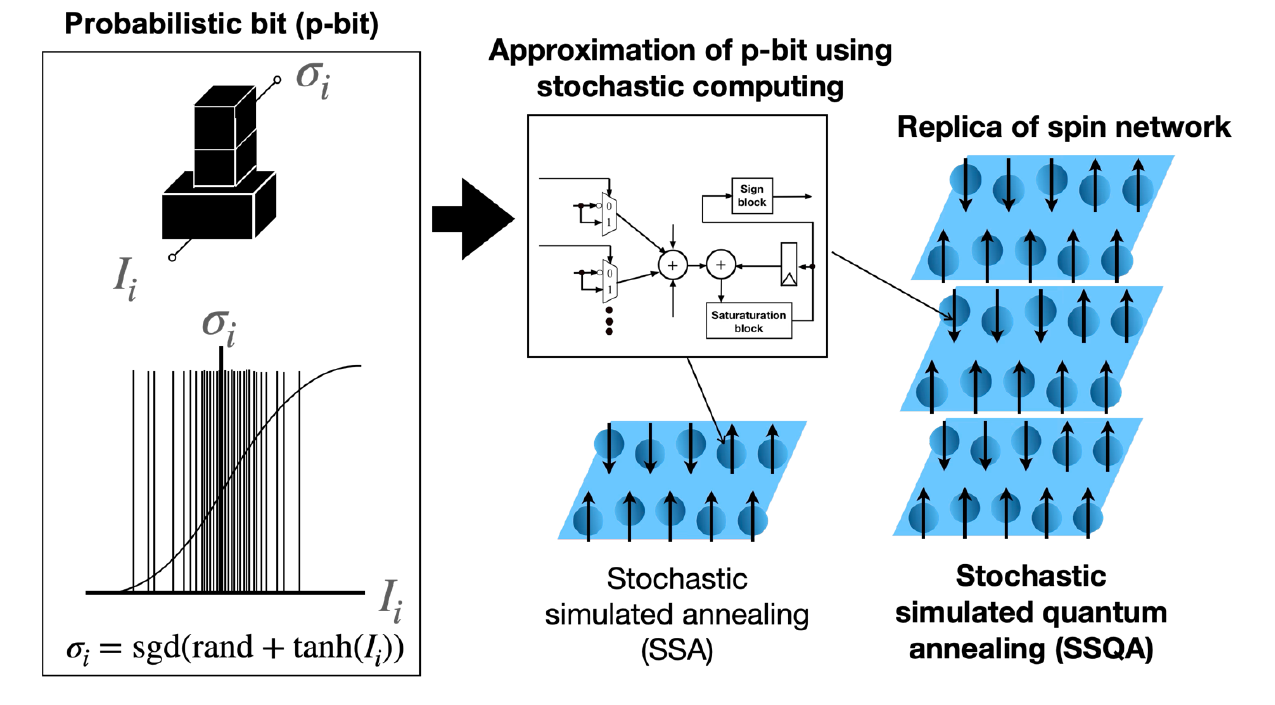}
	\caption{
		Variation of p-bit-based simulated annealing.
		Stochastic simulated annealing (SSA) \cite{SSQA} is a p-bit-based simulated annealing (SA) method approximated using stochastic computing. Stochastic simulated quantum annealing (SSQA) \cite{SSQA} is an alternative p-bit-based SA approach that utilizes replicas of a spin network to mimic quantum annealing on a classical computer.
	}
	\label{fig:pSA}
\end{figure}

A new device model, known as the p-bit, has been proposed \cite{IL}.
This probabilistic nature makes p-bits a valuable tool for solving certain types of problems that require a degree of randomness or uncertainty.
The output state of a p-bit is represented as follows:
\begin{equation}
	\sigma_i(t+1) = {\rm sgn}\Bigl(r_i(t) + {\rm tanh}\bigl(I_i(t+1)\bigr)\Bigr), 
	\label{eqn:pbits}
\end{equation}
where $\sigma_i(t+1) \in \{-1,1\}$ is a binary output signal, $I_i(t+1)$ is a real-valued input signal, and $r_i(t) \in \{-1:1\}$ is a random signal.

p-bit-based simulated annealing (pSA) for Ising models \cite{Ising} has been reported as one of the applications of p-bits \cite{p-bit_general}.
In pSA, each p-bit is biased by $h$ and connected to other p-bits through weights $J$ forming a spin network.
In this context, a p-bit acts as a classical stochastic spin, taking on values of $+1$ or $-1$ with probabilities determined by its input.
This spin network is based on the Ising model, which represents the Hamiltonian (energy function) as follows:
\begin{equation}
	H(\sigma) = - \sum_i h_i\sigma_i - \sum_{i < j} J_{ij}\sigma_i\sigma_j.
	\label{eqn:Ising_SA}
\end{equation}
where $i$ and $j$ ($1 \leq i, j \leq N$) are indices of p-bits, and $N$ is the number of p-bits.
The spins in the Ising spin network are implemented using p-bits.
The input of p-bit $I_i(t+1)$ is calculated using the outputs of other p-bits as defined below:
\begin{equation}
	I_i(t+1) = I_0 \left (h_i+\sum_j J_{ij}\cdot \sigma_j(t) \right),
	\label{eqn:conv}
\end{equation}
where $I_0$ is a pseudo inverse temperature used to control the simulated annealing.

pSA is approximated using stochastic computing, known as stochastic simulated annealing (SSA) \cite{SSQA}, as shown in \cref{fig:pSA}.
Stochastic computing (SC) is a computational paradigm that represents numbers as streams of stochastic bit sequences, allowing arithmetic operations to be performed using simple logic gates \cite{stochastic_first,stochastic}.
Unlike conventional binary computation, SC leverages probability and randomness to achieve low-cost, fault-tolerant, and energy-efficient computing \cite{Sldpc1,Simage,SIIR,SDNN}.
In SSA, \cref{eqn:pbits,eqn:conv} are approximated using SC, which can be implemented in digital circuits.

\subsection{Stochastic simulated quantum annealing (SSQA)}

\begin{figure}[t]
	\centering
	\includegraphics[width=1.0\linewidth]{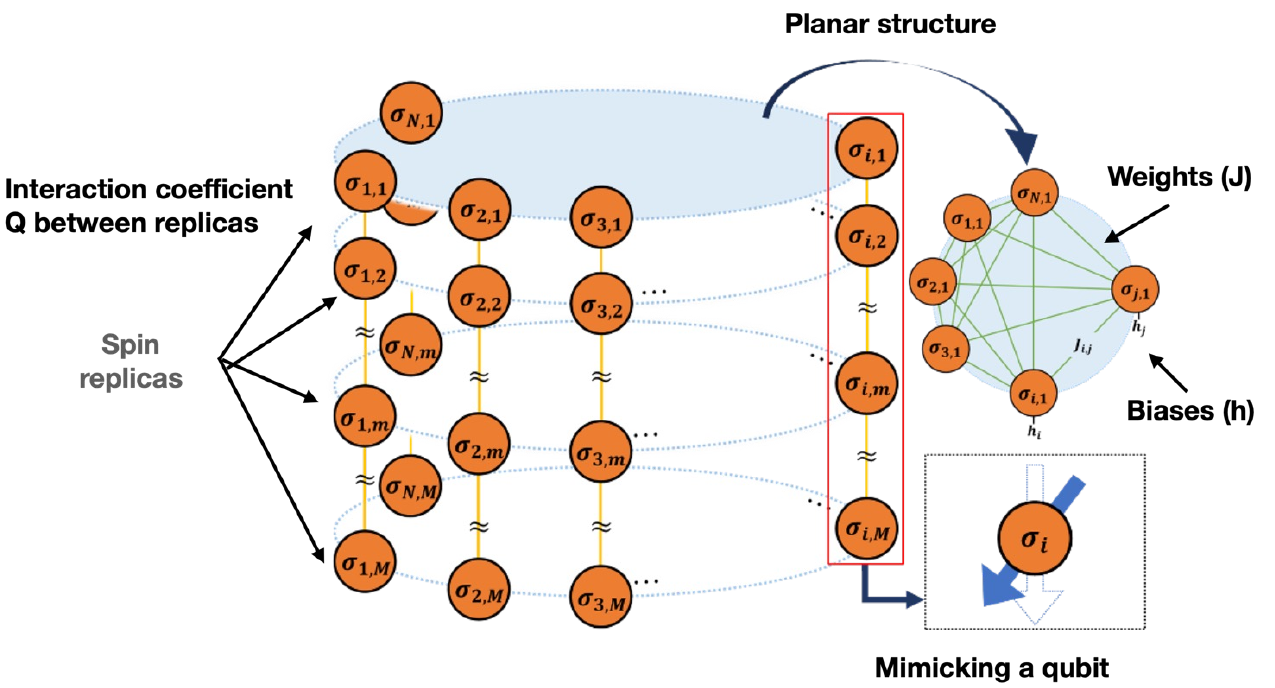}
	\caption{An SSQA spin network consisting of $N \times R$ p-bits, where each replica consists of $N$ p-bits. Each replica of the spin network is based on the Ising model, and adjacent layers are connected through interaction coefficients $Q$.}
	\label{fig:SSQA}
\end{figure}

Stochastic simulated quantum annealing (SSQA) is an alternative p-bit-based SA approach that utilizes replicas of a spin network to mimic quantum annealing on a classical computer \cite{SSQA}.
\cref{fig:SSQA} illustrates a SSQA spin network, which consists of $R$ replicas.
Each replica consists of $N$ p-bits based on the Ising model, and adjacent layers are connected through interaction coefficients $Q$.
The SSQA spin network is represented by a pseudo quantum Hamiltonian, which is approximated by The Trotter-Suzuki decomposition from the quantum Hamiltonian \cite{Suzuki, QMC}. 
This allows its representation using multiple replicas of spin network on classical computers, thereby approximating and representing the time evolution of QA. 
The pseudo quantum Hamiltonian, denoted $H_{c}(\sigma)$, can be expressed as follows:
\begin{equation}
	H_{c}(\sigma) = \sum_{k=1}^{R} \Bigl(H_{p}(\sigma)  
	-  Q\sum_i \sigma_{i,k}\sigma_{i,k+1} \Bigr), 	
	\label{eqn:Ising_QMC}
\end{equation}
\begin{equation}
	H_p(\sigma) = - \sum_i h_i\sigma_{i,k} -  \sum_{i < j} J_{ij}\sigma_{i,k}\sigma_{j,k},
	\label{eqn:Ising_QMC2}
\end{equation}
where $H_p(\sigma)$ is the problem Hamiltonian, $\sigma_{i,k}$ represents the spin state of the $k$-th replica of the spin network ($1 \leq k \leq R$), $R$ is the number of replicas of spins used to represent the q-bit, and $Q$ is a scheduling parameter corresponding to $\Gamma_x$.
The problem Hamiltonian, $H_p(\sigma)$, is the same as the one in  (\cref{eqn:Ising_SA}) used in SA.
%

%
The spin-update algorithm of SSQA is designed based on stochastic computing as SSQA is an extension of SSA.
The update algorithm for the $i$-th spin in the $k$-th replica is as follows:
\begin{subequations}
	\begin{eqnarray}
		I_{i,k}(t+1) &=& h_i+\sum_j J_{ij}\cdot \sigma_{j,k}(t)  + n_{rnd}\cdot r_i(t) \notag \\
		&+& Q(t) \cdot \sigma_{i,k+1}(t-d),
		\label{eqn:I_SC-QMC}
	\end{eqnarray}
	\begin{equation}
		I{s}_{i,k}(t+1)=
		\begin{cases}
			I_0-\alpha, \text{if} \ I{s}_{i,k}(t) + I_{i,k}(t+1) \geq I_0 \\
			-I_0, \text{else if} \ I{s}_{i,k}(t) + I_{i,k}(t+1) < -I_0 \\
			I{s}_{i,k}(t) + I_{i,k}(t+1),  \text{otherwise}
		\end{cases}
		\label{eqn:updown_SC-QMC}
	\end{equation}
	\begin{equation}
		\sigma_{i,k}(t+1)=
		\begin{cases}
			1,& \text{if} \ I{s}_{i,k}(t+1) \geq 0 \\
			-1, & \text{otherwise}.
		\end{cases}
		\label{eqn:m_SC-QMC}
	\end{equation}
	\label{eqn:SC-QMC}
\end{subequations}
where $\sigma_{i,k}(t) \in \{-1,1\}$ and $\sigma_{i,k}(t+1) \in \{-1,1\}$ represent the binary input and output spin states, respectively.
Here, $\sigma_{i,k+1}(t-d)$ represents the $i$-th spin state in the $(k+1)$-th replica, and $d$ is the delay cycle for the coupled effect.
Note that  $d=1$ is used in this paper.
$\alpha$ is a design parameter that defines the offset of the saturation threshold in stochastic computing, and is fixed to 1 throughout the annealing process.
The coupled effect from the upper replica is represented by $Q(t) \cdot \sigma_{i,k+1}(t-d)$.
$I_{i,k}(t+1)$ and $Is_{i,k}(t+1)$ are real-valued internal signals.
In stochastic computing,  \cref{eqn:updown_SC-QMC} and \cref{eqn:m_SC-QMC} approximate $\rm{tanh}(I_0\cdot I_{i,k})$ \cite{stochastic,SDNN}, where $Is_{i,k}$ is the input of \cref{eqn:m_SC-QMC}.
The mean of $\sigma_{i,k}$ is closer to the output of the $\rm{tanh}$ function as stochastic computing is used to approximate $\rm{tanh}$ of the p-bit equation.

\begin{figure}[t]
	\centering
	\includegraphics[width=1.0\linewidth]{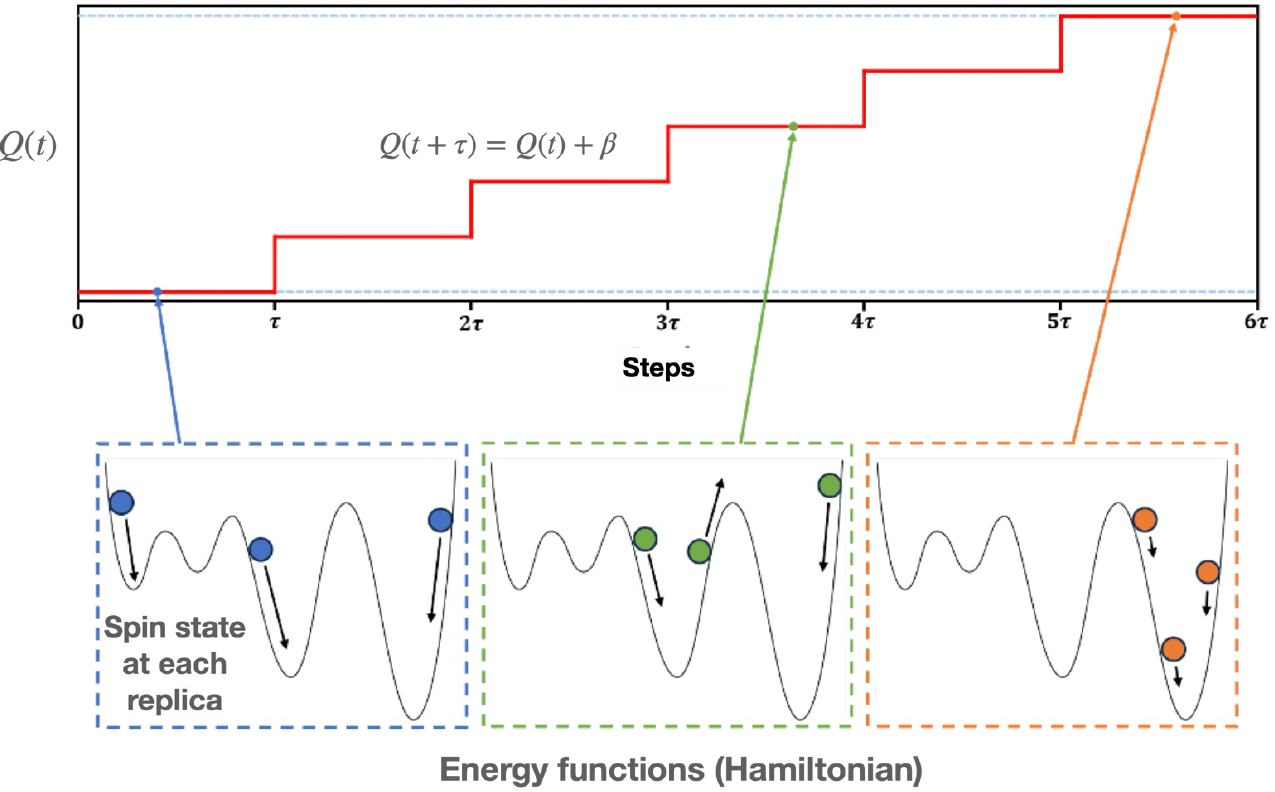}
	\caption{Evolution of the interaction constant $Q(t)$ over the annealing steps in the SSQA method, illustrating how the coupling strength between replicas is gradually increased. The optimization process is guided by this schedule: low $Q(t)$ values allow independent exploration within each replica, while higher $Q(t)$ values enhance inter-replica coupling, encouraging convergence toward the global minimum of the energy function $H$. This mechanism enables efficient solution search via quantum-inspired tunneling behavior.}
	
	\label{fig:Q}
\end{figure}

The SSQA method controls the probability of spin state updates through the interaction constant \( Q(t) \) between networks. 
\cref{fig:Q} represents the interaction constant \( Q(t) \) for the annealing step \( t \) in the SSQA method. 
\( Q(t) \) increases from its minimum value \( Q_{\text{min}} \) to its maximum value \( Q_{\text{max}} \) according to the following equation:
\begin{equation}
	Q(t + \tau) = Q(t) + \beta,
	\label{eqn:Q}
\end{equation}
where \( \tau \) is the number of annealing steps during which the interaction constant remains constant, and \( \beta \) is the increment of the interaction constant. 

When \( Q \) is small, the spin network is loosely coupled to the upper and lower replicas, allowing each replica to explore the minimum value of the problem's energy function independently while minimizing the influence from adjacent replicas. 
On the other hand, when \( Q \) is large, the spin network becomes tightly coupled, enabling it to reach the minimum value by leveraging spin replicas with low energy.

\subsection{Related Work}

Early FPGA realizations of simulated annealing employed spin-parallel datapaths with localized shift registers to cache intermediate states \cite{JETCAS_SSA,Ising_PT}. These designs excel at nearest-neighbor Ising problems but require $O(N)$ multiplexers and incur large routing fan-out, which prevents them from scaling to dense graphs or replica-based algorithms. More recent SSQA prototypes on PYNQ boards \cite{SSQA_FPGA} introduced replica parallelism but retained distributed delay elements, so resource usage still grows with the spin count.

\Cref{tab:related_work} contrasts representative accelerators with the proposed platform. The prior HA-SSA and IPAPT architectures share a spin-parallel organization: HA-SSA leverages stochastic computing to reduce arithmetic complexity, whereas IPAPT instead employs parallel tempering, and both offer limited connectivity and store multiple intermediate replicas, leading to large memory footprints. In contrast, our design centralizes the delay line in dual BRAMs, enabling constant fan-out, single-replica storage, and full-graph programmability.

\begin{table*}[t]
	\centering
	\caption{Feature comparison of representative FPGA annealers.}
	\label{tab:related_work}
	\begin{tabular}{lcccc}
		\toprule
		\textbf{Design} & \textbf{Update topology} & \textbf{Connectivity} & \textbf{Delay storage} & \textbf{Key limitation} \\
		\midrule
		HA-SSA \cite{JETCAS_SSA} & Spin parallel & Four-neighbor lattice & Distributed shift registers & Routing and memory scale with $N$ \\
		IPAPT \cite{Ising_PT} & Spin parallel & Four-neighbor lattice & Distributed shift registers & Replica count tied to logic resources \\
		PYNQ SSQA \cite{SSQA_FPGA} & Replica parallel / spin serial & Sparse graphs & Shift registers per replica & Fan-out and register count grow linearly \\
		\textbf{This work} & Replica parallel / spin serial & Fully connected & Dual BRAM delay line & --- \\
		\bottomrule
	\end{tabular}
\end{table*}

\section{SPIN-SERIAL/REPLICA-PARALLEL HARDWARE ARCHITECTURE FOR SSQA}
\label{sec:hardware}

\subsection{Overall Structure}

\begin{figure}[t]
	\centering
	\includegraphics[width=1.0\linewidth]{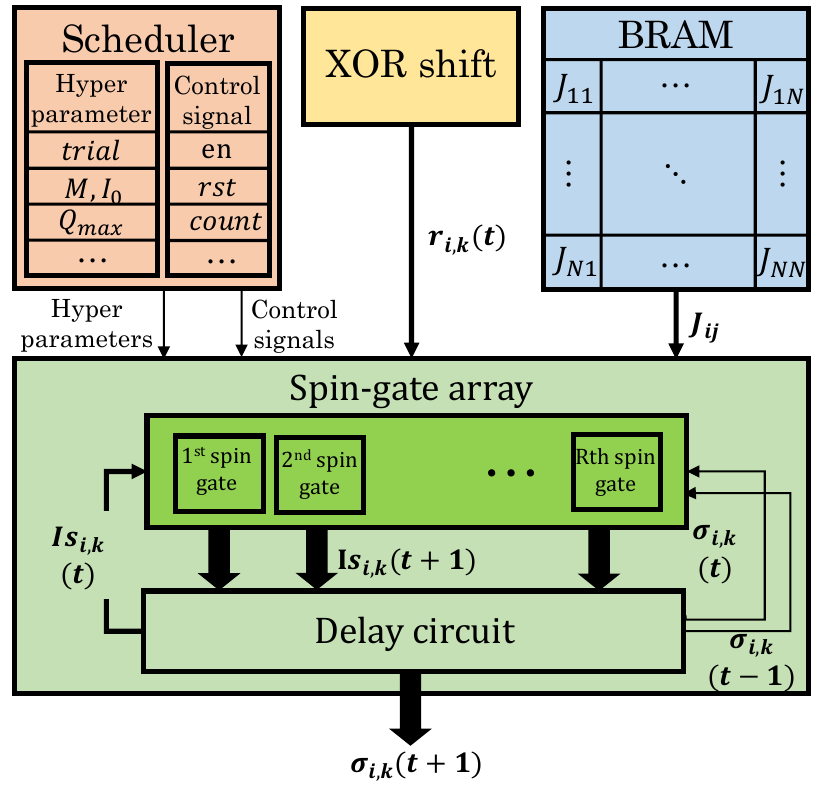}
	\caption{Spin-serial and replica-parallel architecture of the proposed SSQA hardware. The design comprises $R$ spin gate circuits that are reused $N$ times to compute a total of $R \times N$ spins. The spin-serial structure reduces wiring complexity, while the replica-parallel structure allows concurrent access to shared weights, enabling efficient and scalable computation for fully connected graphs.}
	\label{fig:overall}
\end{figure}

\cref{fig:overall} illustrates the proposed hardware architecture designed for implementing the SSQA algorithm described by \cref{eqn:SC-QMC}. The architecture supports a fully connected Ising model comprising $N$ spins and executes computations across $R$ replicas. It employs a spin-serial and replica-parallel approach.
The spin-serial structure significantly reduces the wiring complexity compared to a spin-parallel implementation. 
In a spin-parallel approach, all spin updates are performed simultaneously, requiring each spin gate to access the states of all other spins at once. As a result, each spin gate must incorporate $N$ multiplexers or equivalent routing logic to select and process weighted inputs from the other $N$ spins, leading to increased wiring complexity.
%
%
To minimize hardware complexity, this design adopts serial computation, performing spin interactions sequentially in a time-multiplexed manner.
The replica-parallel structure optimizes memory efficiency, allowing multiple replicas to concurrently access the shared weight matrix $J$. This concurrent access reduces the frequency of memory reads and thus enhances performance.
The proposed architecture consists of the following key components: a spin gate array for computation, BRAM for storing the weight matrix $J$, an XOR-shift random number generator, and a scheduler module responsible for overall control.
Throughout this paper we refer to Xilinx block RAM macros as BRAMs (denoted by the symbol $B$ in the schematics), emphasizing that they are dual-port memories with fixed granularity rather than generic flip-flop arrays.

\begin{figure}[t]
	\centering
	\includegraphics[width=1.0\linewidth]{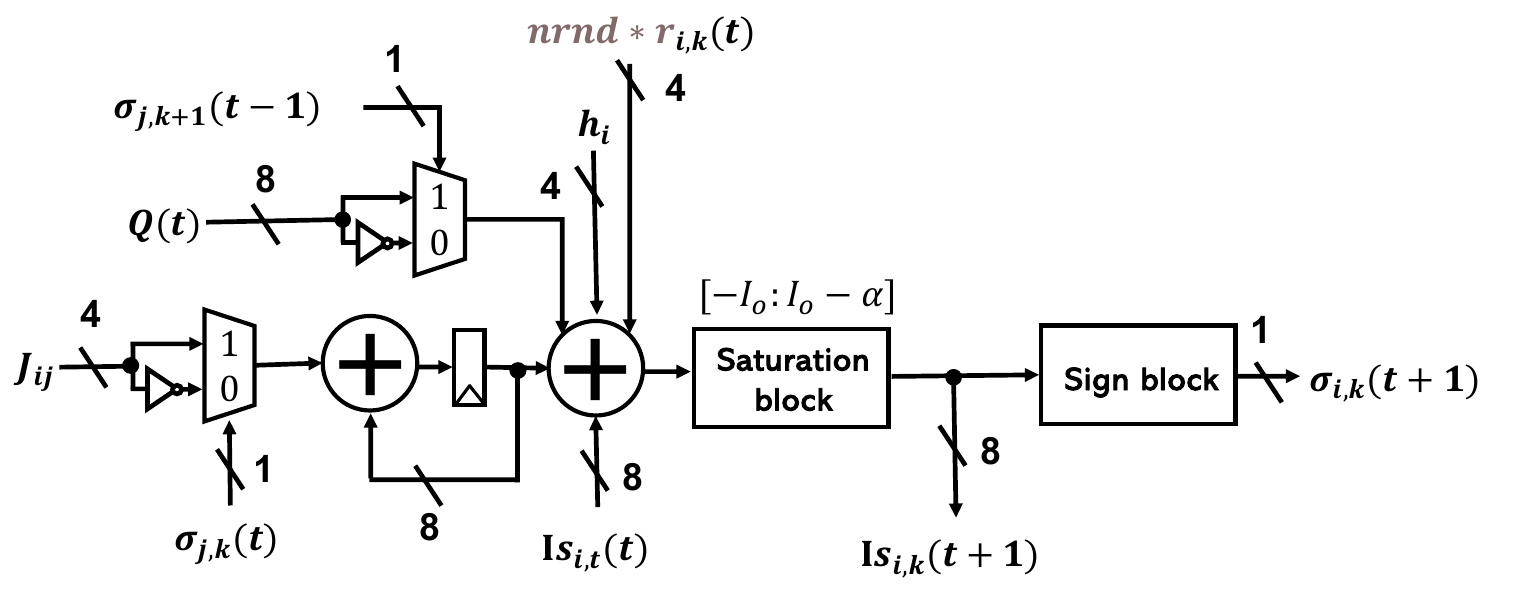}
	\caption{Spin-serial spin gate circuit implementing the update rule in Eq. (6). At each clock cycle, the output of one spin and the corresponding weight $J_{ij}$ are sequentially read from BRAM to compute spin interactions. This time-multiplexed design allows hardware resource reduction while supporting arbitrary spin connectivity.}
	\label{fig:spin_gate}
\end{figure}

The spin gate array comprises $R$ identical spin gate circuits and a delay circuit. \cref{fig:spin_gate} shows the detailed structure of a single spin gate circuit. Following the formulation in \cite{SSQA_FPGA}, each spin gate retains the established stochastic-computing datapath; our contribution is to embed that circuit within a spin-serial scheduler so that a single multiplexer suffices regardless of the number of spin connections.
During each clock cycle, due to the spin-serial design, the output of a specific spin and the associated weight $J_{ij}$ are retrieved from BRAM to compute \cref{eqn:I_SC-QMC}. Repeating this for $N$ clock cycles completes the interaction computations for all connections of one spin. An additional clock cycle computes \cref{eqn:updown_SC-QMC,eqn:m_SC-QMC}, finalizing the update for one spin. This procedure repeats for all $N$ spins, constituting one complete update step as defined by \cref{eqn:SC-QMC}.
As \cref{eqn:I_SC-QMC} requires data from three consecutive time steps ($t+1$, $t$, and $t-1$), a delay circuit retains this necessary data. Two delay circuit implementations are evaluated, with detailed explanations provided in subsequent sections.

Besides spin interactions, the SSQA algorithm requires the generation of various signals, including inverse temperature, random signals and their magnitudes, and interactions between spin networks. 
%
	Random signals $r_{i,k}(t)$ are produced by the XOR-shift random number generator \cite{xorshift}.
	A 64-bit XOR-shift  generator is designed to generate $R$-parallel random signals at each clock cycle.
	The inverse temperature $I_0$ and interaction coefficient $Q(t)$ are generated by the scheduler module.
%
The scheduler module also coordinates the entire circuit operation, including managing input/output processes, enabling, and resetting components. Since hyperparameters influence interaction coefficients and overall control, the scheduler receives these hyperparameters via AXI communication from a CPU integrated into the Zynq FPGA.

\subsection{Shift Register-Based Delay Circuit}

\begin{figure}[t]
	\centering
	\includegraphics[width=1.0\linewidth]{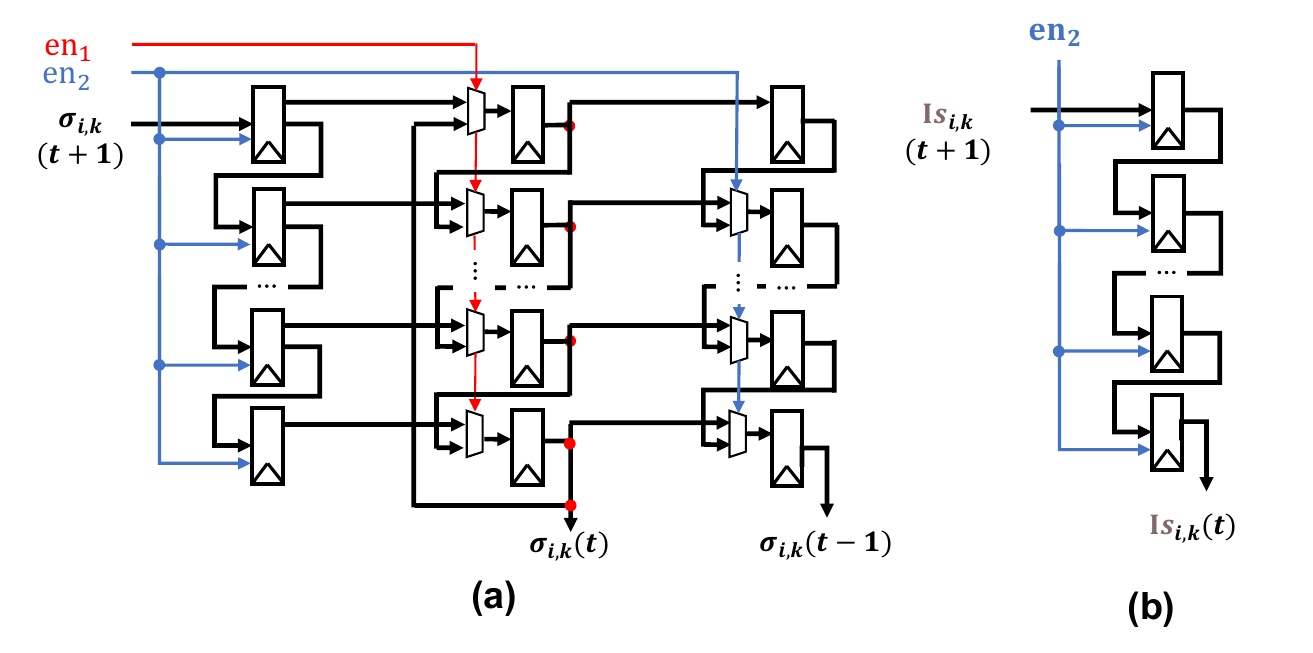}
	\caption{Shift register-based delay circuits used in the $k$-th replicated spin gate: (a) stores spin states $\sigma_{i,k}$ over three consecutive annealing steps, and (b) stores the corresponding saturated internal signals $I_{si,k}$. While straightforward to implement, this design scales poorly with the number of spins due to increased register usage and signal fan-out.}
	\label{fig:shift_register}
\end{figure}

\cref{fig:shift_register}~(a) shows the shift register-based delay circuit used for managing spin states in the $k$-th replicated spin gate~\cite{SSQA_FPGA}. Each spin gate computation requires spin states from three consecutive time steps ($t+1$, $t$, and $t-1$), as described by \cref{eqn:SC-QMC}. Therefore, the delay circuit consists of three sequential register blocks, each containing $N$ registers corresponding to these three time steps.

The operation of each register stage is described as follows:
\begin{itemize}
	\item \textbf{First-stage block} stores newly computed spin states $\sigma_{i,k}(t+1)$ sequentially, shifting data by one register per spin computation. The shift operations are controlled by signals ($en_1$, $en_2$) generated by the scheduler.
	
	\item \textbf{Second-stage block} retains spin states delayed by one annealing step, $\sigma_{i,k}(t)$. After each annealing step, all $N$ spin states computed during the step are simultaneously loaded into this block. During the next annealing step, the block shifts once per spin interaction calculation, providing the appropriate delayed state $\sigma_{i,k}(t)$ sequentially for computation.
	
	\item \textbf{Third-stage block} retains spin states delayed by two annealing steps, $\sigma_{i,k}(t-1)$, and provides them during spin network interaction calculations $Q(t)\cdot\sigma_{i,k+1}(t-1)$. The states from the second-stage block are transferred simultaneously into this block at each annealing step boundary, and subsequently shifted sequentially during interaction calculations.
\end{itemize}

Similarly, \cref{fig:shift_register}~(b) illustrates the delay circuit designed for the saturated internal signals $Is_{i,k}$ within the same replicated spin gate.

While straightforward, the shift register-based delay circuit faces scalability challenges. The circuit requires $3N$ registers, resulting in linear hardware area growth with respect to $N$. Moreover, the fan-out of the control signals also grows with $N$, leading to linear increases in FPGA area due to increased routing complexity.
To overcome these scalability limitations, we propose an alternative dual BRAM-based delay circuit, detailed in the subsequent subsection.

\subsection{Dual BRAM-Based Delay Circuit}

\begin{figure}[t]
	\centering
	\includegraphics[width=1.0\linewidth]{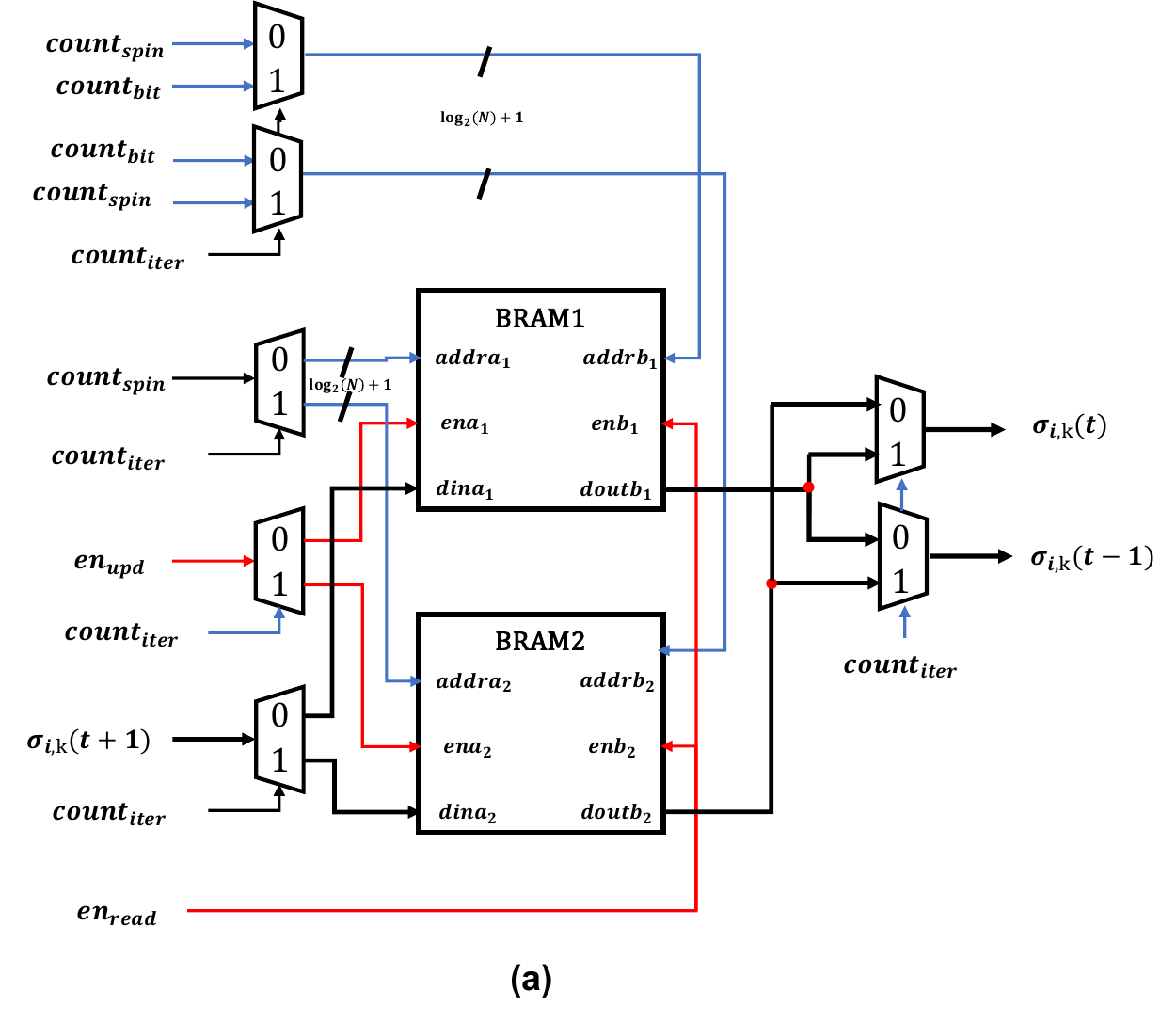}
	\includegraphics[width=1.0\linewidth]{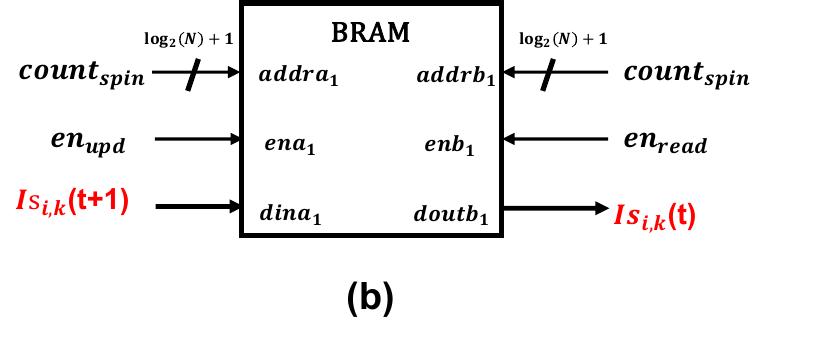}
	\caption{Dual BRAM-based delay circuits for the $k$-th replicated spin gate: (a) handles spin states $\sigma_{i,k}$, and (b) handles saturated internal signals $I_{si,k}$. Two BRAMs alternate read/write operations across annealing steps, enabling delay of one and two time steps. This structure alleviates scalability issues seen in shift register designs by centralizing memory access and minimizing control signal fan-out.}
	\label{fig:BRAM}
\end{figure}

\cref{fig:BRAM}~(a) illustrates the dual BRAM-based delay circuit for storing spin states of the $k$-th replicated spin gate. \cref{fig:BRAM}~(b) depicts a similar delay circuit for saturated internal signals $Is_{i,k}$. Two BRAM blocks alternate operation via a multiplexer at each annealing step, providing delays of one and two annealing steps.
Unlike shift register-based designs, where fan-out problems worsen as the number of spins increases, the proposed dual BRAM-based structure significantly mitigates this issue. BRAM's centralized addressing capability reduces high fan-out wiring, enhancing scalability with increasing spin counts.
At the same time, relying on coarse-grained BRAM macros introduces practical trade-offs: unused capacity leads to fragmentation in small-scale instances, only two memory ports are available per block, and the total replica count is ultimately limited by the number of BRAM tiles on the target FPGA. We therefore co-design the scheduler and memory map to avoid read/write contention and report the resulting utilization in \cref{sec:evaluation}.

The dual BRAM-based delay circuit operates as follows:

\begin{itemize}
	\item During annealing step $t+1$ ($t = 2n$), newly computed spin states $\sigma_{i,k}(t+1)$ are written to BRAM1 using write address \textit{addra1} and write enable signal \textit{ena1}, activated at spin update timing (\textit{enupd}). Conversely, during step $t = 2n - 1$, states are written to BRAM2 using \textit{addra2} and \textit{ena2}.
	
	\item For reading states during step $t$, BRAM1 outputs stored states $\sigma_{i,k}(t)$, accessed via read address \textit{addrb1} corresponding to spin interaction index (\textit{countbit}). During step $t = 2n - 1$, BRAM2 provides $\sigma_{i,k}(t)$ using \textit{addrb2}.
	
	\item At step $t-1$ ($t = 2n$), BRAM1 outputs $\sigma_{i,k}(t-1)$ for replica interaction calculations, with read address set to \textit{countspin}. Similarly, at $t = 2n - 1$, BRAM2 outputs $\sigma_{i,k}(t-1)$ using the same addressing scheme.
\end{itemize}

Since BRAM inherently performs read operations before writes when accessing the same address simultaneously, the integrity of spin state data is preserved during each annealing step.

\section{EVALUATION}
\label{sec:evaluation}

\subsection{Experimental setup}

All software evaluations were executed with an Intel Core-7 7800X (6 cores) and 64~GB of DDR5 memory for the CPU baseline, and an Intel Xeon Gold~6430 (32 cores, AVX-512 enabled), 128~GB of DDR5 memory, and an NVIDIA RTX~4090 GPU for the accelerator experiments. The CPU baseline uses Python~3.6.8 implementations, and the GPU baseline is built with PyCUDA~\cite{PyCUDA}~2022.1 and CUDA~12.2/\texttt{nvcc}~12.2. We fixed the pseudo-random seeds to ensure reproducibility across back-ends.

The G-set benchmarks~\cite{G-set} were used to sweep the number of SSQA replicas $R$ and annealing steps; additional combinatorial optimization instances are introduced in \cref{subsec:apps}. The FPGA design was authored in SystemVerilog, synthesized and placed using Vivado~2023.2 targeting the Xilinx ZC706 board, and validated  for functional equivalence to the software reference. Power estimates rely on Vivado's vector-less analysis with activity factors derived from post-implementation simulation.

\subsection{Algorithm evaluation}

\begin{table}[t]
	\centering
	\caption{Summary of  MAX-CUT problems used for evaluation.}
	\begin{tabular}{c||c|c|c|c|c}
		\hline
		Graph  & \# nodes & Structure & Weights & \# edges  & Best \\
		& & & ($J_{ij}$)  & $(J_{ij}\neq 0)$ & value\\
		\hline
		\hline
		G11  &800  & toroidal & $\{+1, -1\}$  &1600 & 564 \\
		\hline 
		G12 &800 & toroidal  & $\{+1, -1\}$ &1600 & 566\\
		\hline
		G13 & 800  & toroidal & $\{+1,-1\}$ &1600 & 582 \\
		\hline
		G14  & 800  & planar & $\{+1\}$ & 4694 & 3064\\
		\hline
		G15  & 800  & planar & $\{+1\}$ & 4661 & 3050\\
		\hline
	\end{tabular}
	\label{tb:graph}
\end{table}

\begin{figure}[t]
	\centering
	\includegraphics[width=1.0\linewidth]{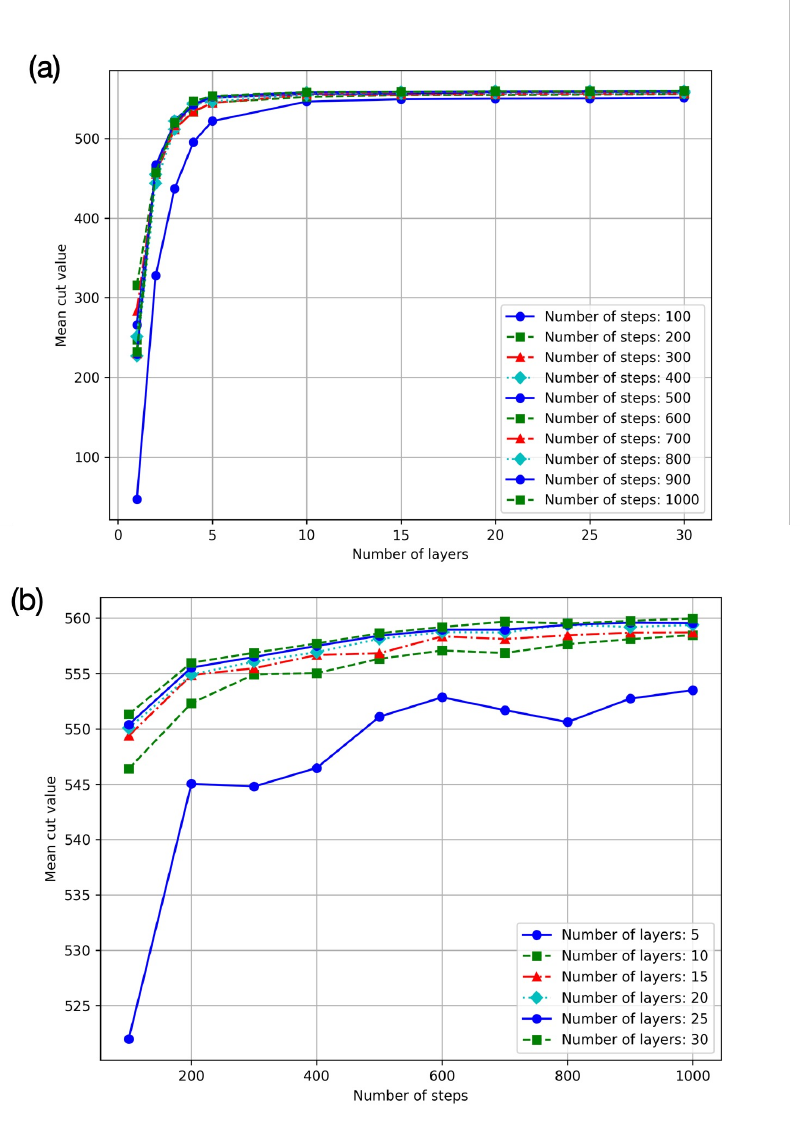}
	\caption{Evaluation of the SSQA algorithm for the G11 instance: (a) average cut value versus the number of layers $R$, showing performance saturation beyond $R = 15$; (b) average cut value versus number of simulation steps for different $R$ values, with saturation observed for $R \geq 20$. 
   } 
	\label{fig:SSQA_param}
\end{figure}

\cref{tb:graph} summarizes the benchmarks for the MAX-CUT problems that are used for evaluation. 
The Gset includes the Gx graphs with different shapes, and weights \cite{G-set}.
\cref{fig:SSQA_param}~(a) illustrates the relationship between the average cut value and the number of layers, $R$, for the SSQA algorithm, focusing on the G11 instance from the G-set. 
Each simulation is conducted according to the following procedure. First, the SSQA algorithm is executed for a predefined number of annealing steps. In this process, $R$ replicas of $N$-spin configurations are evolved in parallel. After the annealing process, the configuration yielding the highest cut value among the $R$ replicas is selected as the final solution.
Each reported value is the average of 100 independent simulation runs, with the number of simulation steps varied from 100 to 1000 in increments of 100. The figure clearly demonstrates that the average cut value saturates beyond 15 layers.
Additionally, \cref{fig:SSQA_param}~(b) presents the relationship between the average cut value and the number of simulation steps for various layer counts ranging from 5 to 30. The results confirm that saturation in performance is achieved with 20 or more layers.

\begin{figure}[t]
	\centering
	\includegraphics[width=1.0\linewidth]{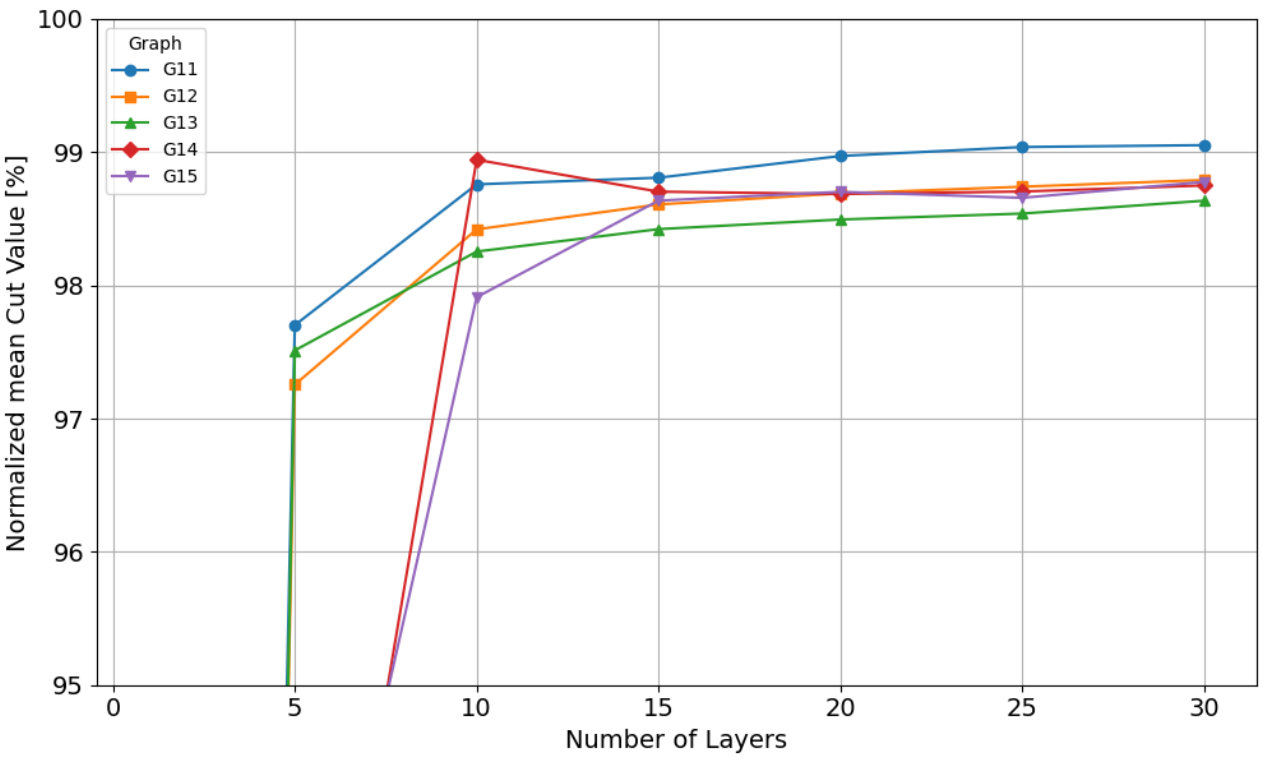}
	\caption{Normalized mean cut value vs. number of layers ($R$) in SSQA.
	Normalized mean cut values nearly saturate  at $R=20$.}
	\label{fig:layer}
\end{figure}

\begin{figure*}[t]
	\centering
	\includegraphics[width=0.9\linewidth]{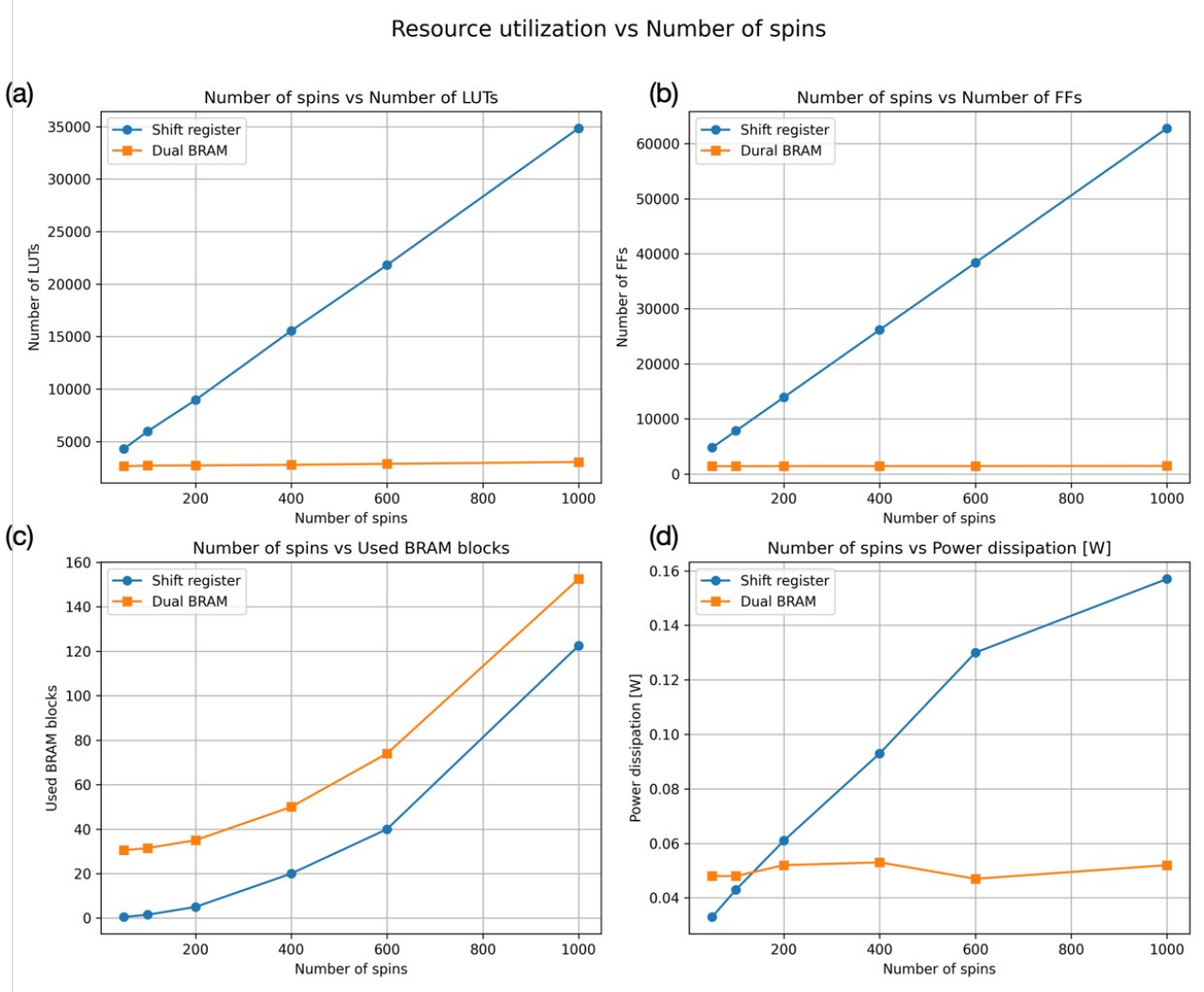}
	\caption{Comparison of resource usage and power consumption between shift register-based and dual-BRAM-based SSQA hardware implementations: (a) LUT usage remains constant in dual-BRAM but grows with spin count in shift-register design; (b) FF usage increases linearly only in the shift-register case; (c) BRAM usage increases with $N^2$, with higher overhead in dual-BRAM due to delay circuit storage; (d) power consumption scales with spin count in the shift-register design but remains stable in dual-BRAM implementation.}
	\label{fig:resource_SSQA}
	\vspace{-3mm}
\end{figure*}

\Cref{fig:layer} extends the replica sweep to all five graphs in \cref{tb:graph}, fixing the annealing steps to 500 while varying $R$ from 1 to 30. Each point aggregates 100 independent trials and is normalized by the best-known cut value. The toroidal (G11--G13) and dense planar (G14--G15) instances all converge within 0.5\% of the optimum once $R \geq 20$, indicating that the replica budget selected on G11 generalizes to both sparse and dense connectivity patterns. Based on these observations we adopt $R=20$ for the hardware implementation, while prior SSQA deployments on graph isomorphism report optimal performance at $R=25$~\cite{SSQA}, highlighting that the best replica count depends on problem structure.

\subsection{Shift register vs. dual BRAM }

We evaluated the SSQA hardware by comparing two types of delay circuits: one based on shift registers \cite{SSQA_FPGA} and the other based on dual BRAMs. Both hardware implementations, each with varying numbers of spins, were designed to operate at a target clock frequency of 100 MHz.
\cref{fig:resource_SSQA}~(a) illustrates the relationship between the number of LUTs and the number of spins. Due to the spin-serial architecture adopted by the proposed SSQA hardware, the LUT count remains nearly constant for the dual-BRAM-based delay circuit, irrespective of the number of spins. Conversely, for the shift-register-based circuit, the LUT count grows proportionally with the number of spins. This increase is primarily due to the expanded fan-out of control signals for the shift registers, necessitating additional buffering elements such as BUFs.
\cref{fig:resource_SSQA}~(b) presents the number of flip-flops (FFs) as a function of the number of spins. For the shift-register-based implementation, the FF count rises linearly with the number of spins, as the size of the delay circuit directly scales with $N$. On the other hand, the dual-BRAM-based delay circuit does not incorporate FFs, thus maintaining an almost constant FF count regardless of spin count.
\cref{fig:resource_SSQA}~(c) shows the number of BRAM blocks relative to the number of spins. The number of BRAM blocks scales proportionally to $N^2$, since the BRAM stores the weight matrix $J$ with size $N^2$. Consequently, the dual-BRAM-based implementation requires more BRAM blocks than the shift-register-based approach, as BRAMs are additionally used within the delay circuit.
\cref{fig:resource_SSQA}~(d) illustrates the power consumption as a function of the number of spins. The shift-register-based implementation's power consumption grows proportionally with spin count due to the linear increase in LUT and FF usage. In contrast, the dual-BRAM-based implementation maintains nearly constant power consumption, independent of the number of spins.

\begin{table}[t]
	\centering
	\caption{Resource utilization of SSQA algorithm on Xilinx ZC706 with 166 MHz for 800 spins.}
	\begin{tabular}{lcc}
		\toprule
		& \textbf{Conventional \cite{SSQA_FPGA}} & \textbf{Proposed} \\
		& (Shift register) & (Dual BRAM) \\
		\midrule
		LUT & 28,525 (13.1\%) & 3,170 (1.45\%) \\
		FF & 50,668 (11.6\%) & 1,643 (0.38\%) \\
		BRAM & 78.5 (14.4\%)& 108.5 (19.9\%)\\
		\midrule
		Power [W] & 0.306 & 0.091 \\
		\bottomrule
	\end{tabular}
	\label{tb:resource}
\end{table}

\cref{tb:resource} summarizes the hardware comparison for \( N = 800 \).
Both implementations were designed and evaluated at a frequency of 166~MHz.
%
The proposed dual-BRAM-based architecture achieved an 89\% reduction in LUT usage, a 97\% reduction in FF usage, and a 70\% reduction in power consumption compared to the shift-register-based design.
%
%
The increase from 78.5 to 108.5 BRAM blocks is a deliberate trade-off: replacing distributed shift registers with two central BRAMs removes the $O(N)$ fan-out without exceeding 20\% of ZC706.

\subsection{Hardware comparisons}

\begin{table}[t]
	\centering
	\caption{Performance comparison of SSQA implementations for 800 spins.}
	\begin{tabular}{lccc}
		\toprule
		& Specification & Clock & Power  \\
		& & frequency & dissipation \\
		\midrule
		\textbf{CPU} & \makecell{Core-7 \\ 7800X} & 3400 MHz & 140 W \\
		\textbf{GPU} & \makecell{NVIDIA \\ RTX4090} & 2235 MHz & 450 W \\
		\textbf{Conventional } & \makecell{Xilinx \\ ZC706} & 166 MHz & 0.306 W \\
		\cite{SSQA_FPGA} \\
		\textbf{Proposed} & \makecell{Xilinx \\ ZC706} & 166 MHz & 0.091 W \\
		\bottomrule
	\end{tabular}
	\label{tb:comp1_transposed}
\end{table}

\begin{figure}[t]
	\centering
	\includegraphics[width=1.0\linewidth]{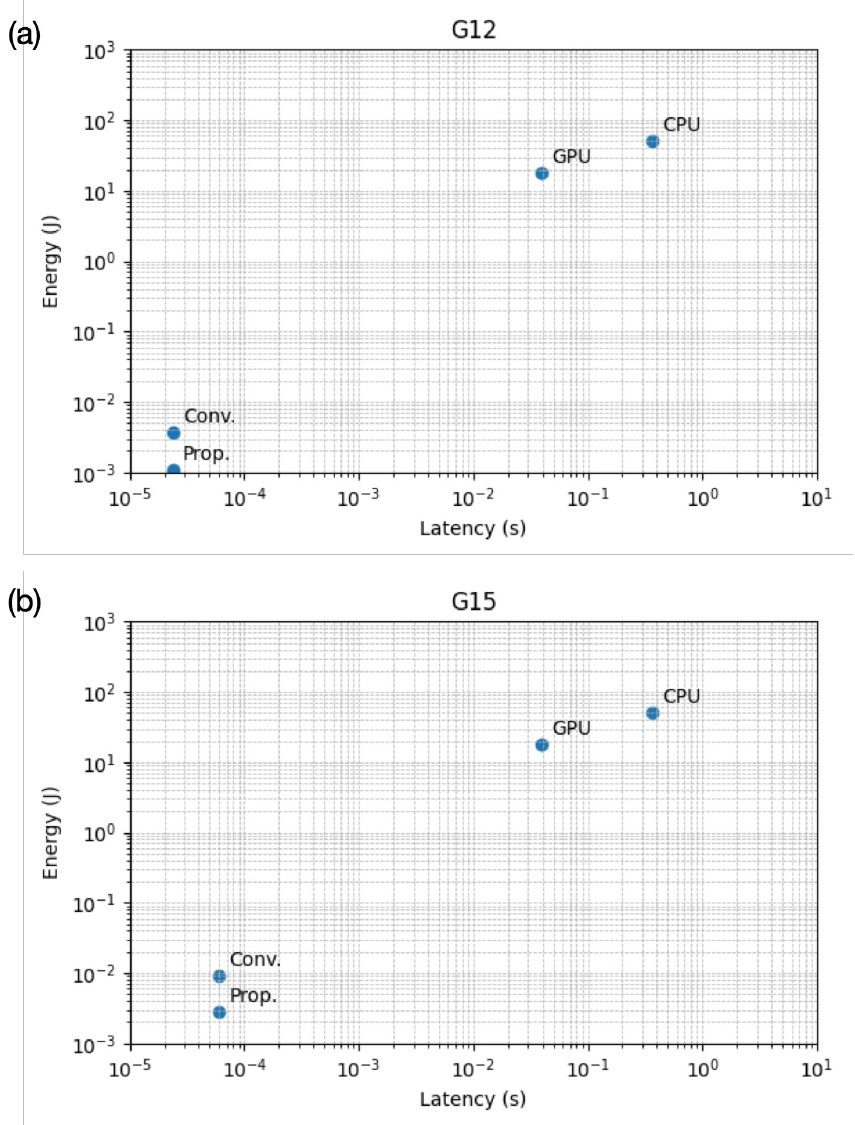}
	\caption{
		Energy–latency trade-off in SSQA implementations for two graph instances (G12 and G15), each executed for 500 annealing steps.  
		(a) and (b) show the results for G12 and G15, respectively.  
		The proposed dual-BRAM-based hardware demonstrates significantly lower energy consumption and latency compared to CPU and GPU implementations.  
		Due to the higher connectivity and larger number of steps in G15, both latency and energy increase for the FPGA implementations.
	}
	
	\label{fig:energy_latency}
\end{figure}

\cref{tb:comp1_transposed} summarizes the performance comparison between the SSQA hardware and its CPU and GPU counterparts.
The latency per annealing step is fixed by the spin-serial schedule: each spin processes the $k$ incident weights and one update, so the total cycles per step are $N\times(k+1)$. For the G-set graphs G11--G13 ($k=4$), one step spans $800\times5$ cycles; fully connected instances with $k=N-1$ scale accordingly. When a graph is sparse, the scheduler bypasses zero-weight placeholders in BRAM, eliminating unnecessary reads and reducing the cycle count while keeping the timing model explicit.

\cref{fig:energy_latency} illustrates the trade-off between energy and latency for SSQA implementations in G12 and G15, with the number of steps fixed at 500.
In G12, the proposed dual-BRAM-based hardware achieves a 97\% reduction in latency and a 99.998\% reduction in energy consumption compared to the CPU implementation.  
When compared with the GPU implementation, it achieves a 70\% latency reduction and a 99.994\% energy reduction.
Due to the increased connectivity in G15 compared to G12, both the latency and energy consumption increase in conventional and proposed hardware implementations.

\begin{table*}[t]
	\centering
	\caption{Performance comparison between a conventional p-bit based hardware~\cite{JETCAS_SSA} and the proposed SSQA-based hardware.}
	\begin{tabular}{c|cc|c|c|cc|c|c}
		\hline
		& \multicolumn{4}{c|}{HA-SSA \cite{JETCAS_SSA}} &  \multicolumn{4}{c}{Proposed} \\
		\hline
	& Best & Average & Memory for & Annealing steps& Best & Average &  Memory for & Annealing steps \\
	Graph & \multicolumn{2}{c|}{cut value} & spin states & & \multicolumn{2}{c|}{cut value} &  spin states &  \\
	\hline
	G11 & 564 & 557 & \multirow{3}{*}{13.2 Mb} & \multirow{3}{*}{90,000} & 564 & 558 & \multirow{3}{*}{32 kb} & \multirow{3}{*}{500} \\
	G12 & 554 & 546 &  &  & 554 & 549 &  &  \\
	G13 & 576 & 570 &  &  & 578 & 573 &  &  \\
	\hline
\end{tabular}
\label{tb:comp2}
\end{table*}

\cref{tab:related_work} and \cref{tb:comp2} together summarize how the proposed SSQA accelerator diverges from prior SSA hardware \cite{JETCAS_SSA}. Because the reference design is spin-parallel and limited to four-neighbor lattices, we benchmark both approaches on the G11--G13 instances that fit its topology, using identical cost functions and annealing schedules wherever possible.
Due to the slow convergence characteristics of SSA compared to SSQA, the number of annealing steps for SSA was set to 90{,}000. In contrast, SSQA achieves comparable or slightly better average cut values using only 500 annealing steps.
Moreover, the conventional hardware requires not only a significantly larger number of annealing steps but also the storage of intermediate spin states, resulting in a large memory footprint of 13.2~Mb BRAM. In contrast, the proposed SSQA hardware only needs to store the final replicas at $R = 20$, reducing the required BRAM size to just 32~kb, which corresponds to a 99.8\% reduction in memory usage.

\begin{table*}[t]
\centering
\caption{Performance comparisons of FPGA implementation for G11.}
\begin{tabular}{lccc}
	\toprule
	& \textbf{Proposed} & \textbf{HA-SSA \cite{JETCAS_SSA}} & \textbf{IPAPT \cite{Ising_PT}} \\
	\midrule
	\textbf{Hardware architecture} & Spin serial & Spin parallel & Spin parallel \\
	\textbf{Graph support} & Fully connected & Limited to 4 neighbors & Limited to 4 neighbors \\
	\textbf{Number of connections per spin} & up to 799 & 4 & 4 \\
	\textbf{Bit width of h and J supported} & 4 & 4 & 2 \\
	\midrule
	\textbf{FPGA} & Xilinx ZC706 & Digilent Genesys 2 & Vertex-5 (XC5VLX330T)\\
	\textbf{Clock frequency} & 166 MHz & 100 MHz & 150 MHz \\
	\textbf{Power dissipation} & 0.091 W & 2.138 W & N/A \\
	\textbf{Annealing latency} & 12.01 ms & 1 ms & 2.64 ms \\
	\textbf{Energy} & 1.093 mJ & 2.138 mJ & N/A \\
	\textbf{Mean cut value} & 558.4 & 558 & 561 \\
	\midrule
	\textbf{LUT} & 3,170 (1.45 \%)& 105,294 (51.7 \%) & 46,753 (22.5 \%)\\
	\textbf{FF} & 1,643 (0.38 \%)& 13,692 (3.36 \%)& 19,797 (9.55 \%) \\
	\textbf{BRAM blocks} & 108.5 (19.9 \%)& 356 (79.9 \%) & N/A \\
	\bottomrule
\end{tabular}
\label{tb:comp3}
\end{table*}

\cref{tb:comp3} presents a performance comparison between the proposed method and conventional annealing architectures from previous studies, using the G11 benchmark as the target graph. Both conventional methods adopt a spin-parallel architecture, which limits their applicability to graphs with only nearest-neighbor connections.
In contrast, the proposed architecture employs a spin-serial structure, enabling the processing of fully connected graphs. Due to this architectural difference, the annealing latency of the proposed method is 12~ms, which is higher than that of the two conventional approaches.
However, the proposed method achieves significant hardware resource savings. Compared to the HA-SSA architecture \cite{JETCAS_SSA}, it reduces energy consumption by 50\%, the number of LUTs by 97\%, the number of FFs by 88\%, and the number of BRAM blocks by 70\%. Compared to the IPAPT architecture \cite{Ising_PT}, the number of LUTs and FFs are reduced by 93\% and 92\%, respectively.

\begin{figure}[t]
	\centering
	\includegraphics[width=1.0\linewidth]{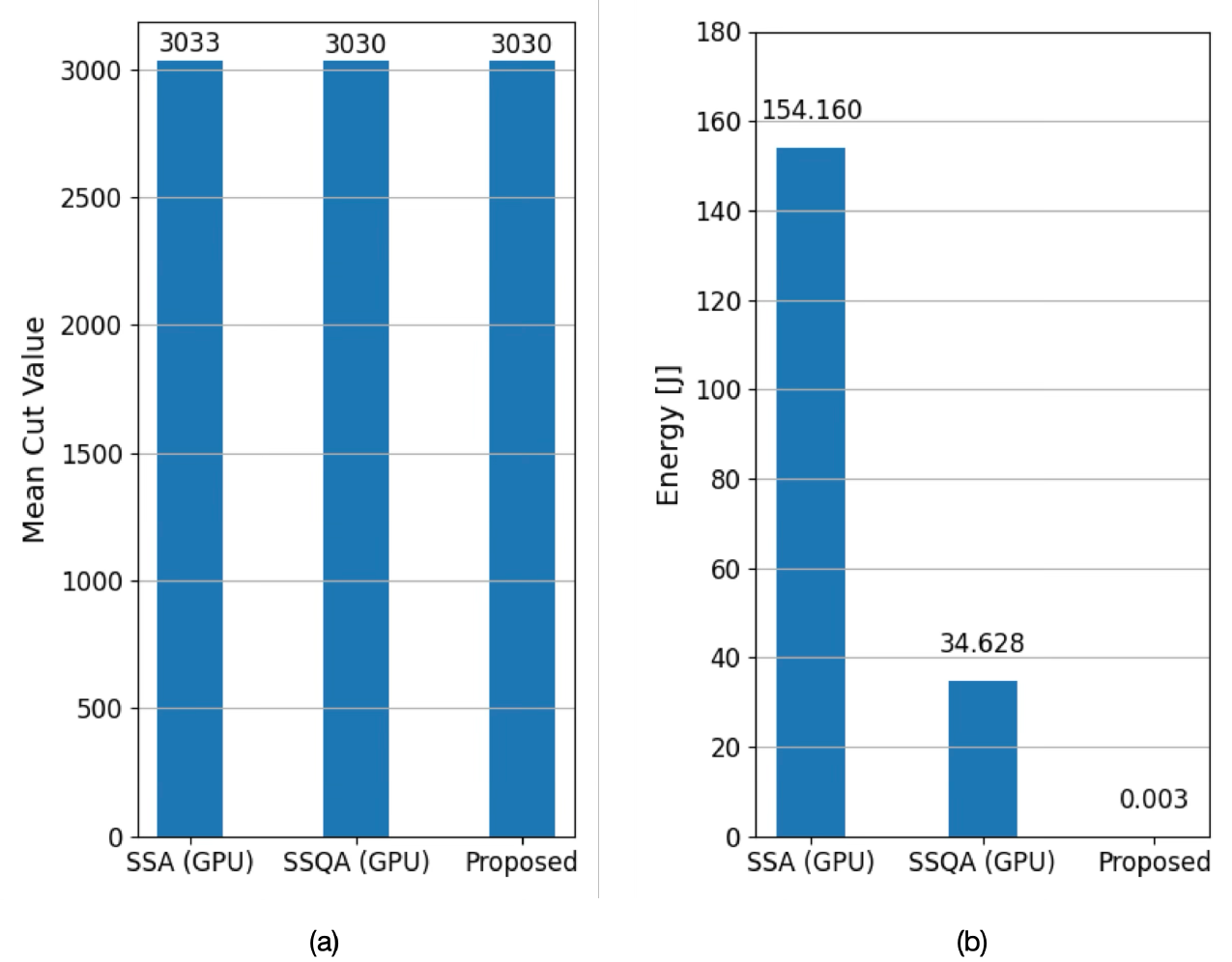}
	\caption{
		Comparisons of mean cut value and energy for G14:
		(a) mean cut values obtained for 100 trials and 
		(b) annealing energy.
	}
	\label{fig:G14}
\end{figure}

\cref{fig:G14} shows the mean cut values and annealing energy for the MAX-CUT benchmark G14.
Conventional hardware architectures were limited to processing graphs with only four adjacent nodes. As a result, evaluations were restricted to benchmarks such as G11, G12, and G13 \cite{Ising_PT,JETCAS_SSA}.
In contrast, the proposed hardware is capable of handling fully connected graphs, which allows it to process more complex topologies like G14, where each node is connected to more than four others.
Since conventional hardware cannot process such dense graphs, we conducted a performance comparison between the proposed SSQA algorithm implemented on GPU and the conventional SSA algorithm on GPU instead.
The SSA algorithm runs for 10,000 annealing steps and the SSQA algorithms with $R=20$ run for 500 steps.
Compared to SSA (GPU) and SSQA (GPU), the proposed method achieved comparable average cut values while reducing the energy consumption by 99.998\% and 99.992\%, respectively.

\section{SENSITIVITY ANALYSES AND DISCUSSION}
\label{sec:discussion}

\subsection{Latency-area trade-off}
Owing to the node-serial spin update, the present implementation
requires 12.0\,ms to solve the 800-spin G11 instance—longer than the
1 ms of the previous HA-SSA FPGA.
Nevertheless, our design consumes only 20\,\% of the XC7Z045 fabric,
yielding an \emph{area–delay product (ADP)} of
\(\text{ADP}_{\text{serial}} = 0.199 \times 12.0\text{\,ms} = 2.39\text{\,ms}\).
Note that \text{area} is $\max\{\text{LUT\%},\text{FF\%},\text{BRAM\%}$.
In the proposed hardware, $A=19.9\%$ (BRAM-dominated).
For edge-class optimisers, an interaction latency of 50 ms is
generally accepted \cite{martens2018latency}; hence the serial
configuration already meets real-time constraints while
cutting area—and thus cost and idle power—by more than 60 \% compared
with HA-SSA.

Because the datapath is fully pipelined, latency can be
\emph{linearly reduced} by instantiating $p$ parallel spin engines.
A ten-way parallel variant, for example, shortens latency to
1.2 ms while lifting resource utilisation to 54.8 \%, resulting in
\(\text{ADP}_{p=10}=0.648 \text{\,ms}\) (3.7 × lower than the serial design) and
still remaining below the 80 \% BRAM ceiling of HA-SSA.
The constant energy per solve (1.1 mJ) stems from the proportional
increase in power with~$p$, implying that designers may trade silicon
for speed without compromising battery life.

Further latency reductions could be achieved by
(i) selectively parallelising high-degree spins,
(ii) interleaving replica updates to hide memory latency,
and (iii) sparsifying or quantising the weight matrix so that
the BRAM footprint scales sub-linearly with problem size.
Compression schemes such as run-length or delta encoding would
release additional BRAM blocks, enabling graphs well beyond
10\,000 spins to fit on mid-range FPGAs.
These enhancements will broaden the applicability of the architecture
to real-time machine-learning inference, embedded optimisation, and
event-driven neuromorphic workloads.

Across the sweep of $N=100$ to $N=800$ spins in \cref{fig:resource_SSQA}, LUT and FF usage vary by less than 5\%, confirming that the dual-BRAM delay line decouples logic cost from graph density. Power measurements follow the same trend, underscoring that scalability derives from the architectural changes rather than device-specific tuning.

\subsection{Applicability to other combinatorial optimization problems}
\label{subsec:apps}
Beyond MAX-CUT, the SSQA algorithm itself has been validated on other combinatorial benchmarks. We apply the same replica-coupled update rule to traveling salesman (TSP) and graph isomorphism (GI) instances, demonstrating algorithm-level convergence on both domains~\cite{SSQA}. In particular, their GI study shows that with $R=25$ replicas SSQA sustains a 51\% success probability at $N=2{,}025$ nodes and cuts the time-to-solution (TTS) to 146~s, a 91.4\% reduction relative to SSA's 1{,}690~s TTS under the same cycle budget; SA requires 62{,}022~s for the same instance (423$\times$ slower). Even for $N=2{,}500$, SSQA still converges in 405~s with 41\% success while SSA fails to reach the optimum~\cite{SSQA}. Because our FPGA bitstream implements the identical update rules and replica interactions, those GI and TSP instances---and any problem that admits an equivalent QUBO formulation~\cite{Ising}---can be executed by updating only the BRAM initialization files, without architectural changes.

The synthesizable SystemVerilog source code, host-side drivers, and Python reference implementation are available in our public repository~\cite{nonizawaSSQA2024GitHub}. The project includes Vivado~2023.2 configuration scripts, BRAM initialization files, and unit tests that mirror the benchmarks discussed above to aid reproducibility.

\begin{table*}[t]
\centering
\caption{Qualitative comparison of FPGA-based annealing architectures}
\label{tab:qualitative_comparison}
\begin{tabular}{lcccc}
	\toprule
	& \cite{yoshimura2017implementation} & \cite{yoshioka2023fpga} & \cite{waidyasooriya2021highly} & This work \\
	\midrule
	Hardware cost (LUTs/FFs) & Small & Large & Large & Small \\
	Graph configuration       & 2D nearest-neighbor & Fully connected & Fully connected & Fully connected \\
	Scheduling logic          & Complex & Simple & Simple & Simple \\
	Power consumption         & Low & High & High & Low \\
	Processing speed          & High & High & Low & Middle \\
	Energy efficiency         & High & Low & Low & High \\
	\bottomrule
\end{tabular}
\end{table*}

\subsection{Comparison with Ising/QUBO annealers}

\cref{tb:comp3} focuses on FPGA-class baselines, yet fully connected Ising/QUBO accelerators also exist in custom CMOS and quantum hardware. Fujitsu's Digital Annealer~\cite{DA} implements a massively parallel simulated annealer that solves dense QUBOs of up to 8{,}192 variables but requires datacenter-scale power budgets (>100~W) and proprietary cooling. Superconducting quantum annealers such as D-Wave Advantage~\cite{Boixo2014} offer quantum tunneling and larger solution spaces but operate at milli-Kelvin temperatures and impose sparse Chimera/Pegasus connectivity, necessitating costly minor-embedding. Our FPGA-based p-bit engine instead targets sub-watt envelopes and natively supports arbitrary connectivity without embedding overhead.

As summarized in \cref{tab:qualitative_comparison}, among FPGA implementations the proposed architecture provides the lowest LUT/FF footprint while maintaining full connectivity. Resource-sharing~\cite{yoshimura2017implementation} and chaotic Boltzmann machine~\cite{yoshioka2023fpga} designs either forego dense graphs or sacrifice energy efficiency; the highly-parallel SQA system~\cite{waidyasooriya2021highly} attains massive scale but pays a 5--270~ms step latency (roughly $2.1\times10^{2}$–$1.1\times10^{4}$ slower than our 24~$\mu$s), even before considering power.

The contrast also clarifies the advantages and trade-offs of p-bit annealers. Unlike analog Ising machines that rely on continuous device dynamics, p-bit fabrics store probabilistic states digitally, making them resilient to fabrication variability and amenable to formal verification. Compared with QUBO-centric digital annealers, our architecture consumes less memory by streaming weights and storing only final replicas, at the expense of higher per-step latency. These properties make the proposed design attractive for embedded optimisation workloads where energy and programmability outweigh absolute throughput.

\section{Conclusion}
\label{sec:conclusion}

We presented an energy-efficient hardware architecture for fully connected simulated annealing, combining a quantum-inspired SSQA algorithm with a spin-serial/replica-parallel design and a dual-BRAM delay scheme. This architecture addresses both connectivity and scalability challenges in conventional p-bit-based annealing hardware. Unlike nearest-neighbor-limited or shift-register-based approaches, our design supports fully connected graphs while maintaining nearly constant LUT and FF usage as the number of spins increases, thanks to the separation of logic and memory enabled by the dual-BRAM scheme.
Evaluation on the G11 instance from the G-set benchmark confirmed significant resource and energy efficiency, achieving over 50\% reduction in energy consumption, 97\% reduction in logic resource usage, and a 70\% reduction in BRAM consumption, without compromising solution quality. These results demonstrate the feasibility of scalable, quantum-inspired optimization hardware on modern FPGAs.

Future work includes reducing BRAM usage through sparsification, quantization, or compression of the weight matrix, as well as extending support to other combinatorial problems such as graph coloring and the traveling salesman problem. Additionally, integration with emerging memory devices such as MRAM or spintronic-based p-bits will be explored to further improve efficiency and practicality for real-world applications.

\bibliographystyle{IEEEtran}
\bibliography{SSQA}
	
\end{document}